\documentstyle[12pt,epsf]{article} 
\begin{document}            

\title{Spin-Peierls instability 
       \protect\\
       in the spin-$\frac{1}{2}$ transverse $XX$ chain
       \protect\\
       with Dzyaloshinskii-Moriya interaction}

\author{O. Derzhko$^{a,b}$, J. Richter$^{c}$ and O. Zaburannyi$^{a}$\\
\small{$^{a}$Institute for Condensed Matter Physics,}\\
\small{1 Svientsitskii St., L'viv-11, 290011, Ukraine}\\
\small{$^{b}$Chair of Theoretical Physics, 
Ivan Franko State University of L'viv,}\\
\small{12 Drahomanov St., L'viv-5, 290005, Ukraine}\\
\small{$^{c}$Institut f\"{u}r Theoretische Physik,
Universit\"{a}t Magdeburg,}\\
\small{P.O. Box 4120, D-39016 Magdeburg, Germany}}

\date{\today}

\maketitle                   

\begin{abstract}
We calculate exactly the density of magnon states
of the regularly alternating spin-$\frac{1}{2}$ $XX$ chain
with Dzyaloshinskii-Moriya interaction.
The obtained result permit us to examine the stability of the chain
with respect to spin-Peierls dimerization.
We found that 
depending on the dependences of
Dzyaloshinskii-Moriya interaction 
on distortion amplitude
it
may act 
either in favour of the dimerization
or against the dimerization.
\end{abstract}

\vspace{5mm}


\vspace{5mm}


\vspace{5mm}

\noindent
{\bf {Postal addresses:}}\\

\vspace{0mm}

\noindent
Dr. Oleg Derzhko (corresponding author)\\
Oles' Zaburannyi\\
Institute for Condensed Matter Physics\\
1 Svientsitskii Street, L'viv-11, 290011, Ukraine\\
tel/fax: (0322) 76 19 78\\
email: derzhko@icmp.lviv.ua\\

\vspace{0mm}

\noindent
Prof. Johannes Richter\\
Institut f\"{u}r Theoretische Physik,
Universit\"{a}t Magdeburg\\
P.O. Box 4120, D-39016 Magdeburg, Germany\\
tel: (0049) 391 671 8841\\
fax: (0049) 391 671 1217\\
email: Johannes.Richter@Physik.Uni-Magdeburg.DE

\clearpage

\renewcommand\baselinestretch{1.7}
\large\normalsize

The discovery of the inorganic spin-Peierls compound
CuGeO$_3$
\cite{001,002}
renewed an interest in the investigation 
of spin-Peierls instability of quantum spin chains.
Up to date  a large number of papers concerning the quantum spin chains 
which are believed to model appropriately the spin degrees of freedom of 
the spin-Peierls compounds has appeared. As a rule the considered models 
are rather complicated and have been examined with exploiting different 
approximations. On the other hand, some generic features of the spin-Peierls 
systems can be illustrated in the simplified but exactly solvable models. As 
an example of such model one can refer to the spin-$\frac{1}{2}$ $XX$ chain
that was studied in several papers
\cite{003,004,005}.
The purpose of the present paper is to examine the influence
of introducing Dzyaloshinskii-Moriya interspin coupling
on the
spin-Peierls dimerization within the frames of the one-dimensional 
spin-$\frac{1}{2}$ $XX$ model in a transverse field.
The presence of Dzyaloshinskii-Moriya interaction  
for CuGeO$_3$ was proposed
in Refs. \cite{006,007,008}
in order to explain the EPR and ESR experimental data.
On the other hand,
the multisublattice spin-$\frac{1}{2}$ $XX$ chain
with Dzyaloshinskii-Moriya interaction was introduced in
\cite{009},
however, that paper was not devoted to the study of spin-Peierls 
instability.
In our paper we closely follow the idea of
Ref. \cite{003}
and compare the total ground state energy
of the dimerized and uniform chains.
In contrast to previous works
\cite{003,004,005,009}
we use the continued-fraction
representation for one-fermion Green functions
\cite{010,011,012,013}
that allows one
to consider in a similar way not only the dimerized lattice but
also more complicated lattice distortions.
Based on the performed calculations we found that Dzyaloshinskii-Moriya 
interaction may act both in favour of the dimerization and against the 
dimerization. The result of its influence depends on the dependence 
of Dzyaloshinskii-Moriya interaction on the distortion amplitude in 
comparison with such a dependence of the isotropic exchange interaction.

We consider $N\rightarrow \infty$ spins $\frac{1}{2}$ on a circle
with the Hamiltonian
\begin{eqnarray}
\label{001}
H
=2\sum_{n=1}^NI_n\left(s_n^xs_{n+1}^x+s_n^ys_{n+1}^y\right)
\nonumber\\
+2\sum_{n=1}^ND_n\left(s_n^xs_{n+1}^y-s_n^ys_{n+1}^x\right)
\nonumber\\
+\sum_{n=1}^N\Omega_ns_n^z.
\end{eqnarray}
Here $I_n$ and $D_n$ are the isotropic exchange coupling and  
Dzyaloshinskii-Moriya coupling between the neighbouring sites $n$ and $n+1$ 
and $\Omega_n$ is the transverse field at site $n$.
We restricted ourselves to the Hamiltonian (\ref{001})
since
after the Jordan-Wigner transformation 
it reduces to tight-binding
spinless fer\-mi\-ons.
We introduce the temperature double-time
one-fermion Green functions that yield the density of magnon states
by the relation
$\rho(E)
=\mp\frac{1}{\pi N}\sum_{n=1}^N{\mbox{Im}}G_{nn}^{\mp}$,
$G_{nm}^{\mp}\equiv G_{nm}^{\mp}(E\pm i\epsilon)$.
In the case of the tight-binding fermions the required diagonal Green 
functions can be expressed by means of continued fractions 
\cite{010,011,012,013}
\begin{eqnarray}
\label{002}
G_{nn}^{\mp}=\frac{1}{E\pm i\epsilon-\Omega_n-\Delta_n^--\Delta_n^+},
\\
\Delta_n^-=\frac{I^2_{n-1}+D^2_{n-1}}
{E\pm i\epsilon -\Omega_{n-1}-\frac{I^2_{n-2}+D^2_{n-2}}
{E\pm i\epsilon -\Omega_{n-2}-_{\ddots}}},
\nonumber\\
\Delta_n^+=\frac{I^2_{n}+D^2_{n}}
{E\pm i\epsilon -\Omega_{n+1}-\frac{I^2_{n+1}+D^2_{n+1}}
{E\pm i\epsilon -\Omega_{n+2}-_{\ddots}}}.
\nonumber
\end{eqnarray}
For any periodic configuration of the intersite couplings and
transverse field the fractions
$\Delta^-_n$ and $\Delta^+_n$
involved into $G^{\mp}_{nn}$ (\ref{002})
become finite
and can be evaluated exactly
yielding obviously the exact result for the density of states
$\rho(E)$
and hence for the thermodynamic quantities of spin model (\ref{001}).

In what follows we shall use the 
result for the periodic chain having period 2
that is characterized by a sequence 
$I_1D_1\Omega_1I_2D_2\Omega_2I_1D_1\Omega_1I_2D_2\Omega_2\ldots\;$.
For such a case we have
\begin{eqnarray}
\label{003}
\rho(E)=
\left\{
\begin{array}{ll}
0, &
{\mbox{if}}\;\;\;E\le b_4,\;b_3\le E\le b_2,\;b_1\le E,\\
\frac{1}{2\pi}\frac{\vert 2E-\Omega_1-\Omega_2\vert}
{\sqrt{{\cal{B}}(E)}},&
{\mbox{if}}\;\;\;b_4<E<b_3,\;b_2<E<b_1,
\end{array}
\right.
\\
{\cal{B}}(E)
=4{\cal{I}}_1^2{\cal{I}}_2^2
-\left[
(E-\Omega_1)(E-\Omega_2)-{\cal{I}}_1^2-{\cal{I}}_2^2
\right]^2
\nonumber\\
=-(E-b_4)(E-b_3)(E-b_2)(E-b_1),
\nonumber\\
\left\{b_4\le b_3\le b_2\le b_1\right\}
=
\left\{
\frac{1}{2}\left(\Omega_1+\Omega_2\right)\pm {\sf{b}}_1,
\frac{1}{2}\left(\Omega_1+\Omega_2\right)\pm {\sf{b}}_2
\right\},
\nonumber\\
{\sf{b}}_1=\frac{1}{2}
\sqrt{\left(\Omega_1-\Omega_2\right)^2
+4\left(\vert{\cal{I}}_1\vert+\vert{\cal{I}}_2\vert\right)^2},
\nonumber\\
{\sf{b}}_2=\frac{1}{2}
\sqrt{\left(\Omega_1-\Omega_2\right)^2
+4\left(\vert{\cal{I}}_1\vert-\vert{\cal{I}}_2\vert\right)^2},
\nonumber\\
{\cal{I}}^2_n=I_n^2+D_n^2.
\nonumber
\end{eqnarray}

Let us examine the instability
of the considered spin chain
with respect to dimerization. 
To do this we assume
$\vert I_1\vert=\vert I\vert(1+\delta)$,
$\vert I_2\vert=\vert I\vert(1-\delta)$,
$\vert D_1\vert=\vert D\vert(1+k\delta)$,
$\vert D_2\vert=\vert D\vert(1-k\delta)$,
where
$0\le\delta\le 1$
is the dimerization parameter.
Putting $k=0$ one has a chain in which Dzyaloshinskii-Moriya interaction 
does not depend on the lattice distortion, whereas for $k=1$ the dependence 
of Dzyaloshinskii-Moriya interaction on the lattice distortion is as 
that for the isotropic exchange interaction. 
We consider a case of zero temperature and look for the 
total energy per site
${\cal{E}}(\delta)$
that consists of the magnetic part 
$e_0(\delta)$
and the elastic part
$\alpha\delta^2$. 
From (\ref{003}) one finds that
\begin{eqnarray}
\label{004}
e_0(\delta)
=-\frac{1}{2}\int_{-\infty}^{\infty}
dE\rho(E)\vert E\vert
\nonumber\\
=-\frac{1}{\pi}{\sf{b}}_1
{\mbox{E}}\left(\psi,\frac{{\sf{b}}_1^2-{\sf{b}}_2^2}{{\sf{b}}_1^2}\right)
-\frac{1}{2}
\left\vert\Omega_1+\Omega_2\right\vert
\left(
\frac{1}{2}-\frac{\psi}{\pi}
\right),
\\
\psi
=\left\{
\begin{array}{ll}
0 & {\mbox{if}} \;\;\; 
{\sf{b}}_1\le\frac{1}{2}\vert\Omega_1+\Omega_2\vert,\\
{\mbox{arcsin}}\sqrt{
\frac{{\sf{b}}_1^2-\frac{1}{4}\left(\Omega_1+\Omega_2\right)^2}
{{\sf{b}}_1^2-{\sf{b}}_2^2}} &
{\mbox{if}} \;\;\; 
{\sf{b}}_2\le\frac{1}{2}\vert\Omega_1+\Omega_2\vert
<{\sf{b}}_1,\\
\frac{\pi}{2} &  {\mbox{if}} \;\;\;  
\frac{1}{2}\vert\Omega_1+\Omega_2\vert<{\sf{b}}_2,
\end{array}
\right.
\nonumber\\
{\sf{b}}_1=
\frac{1}{2}
\sqrt{\left(\Omega_1-\Omega_2\right)^2
+4
\left[
\sqrt{I^2(1+\delta)^2+D^2(1+k\delta)^2}
+\sqrt{I^2(1-\delta)^2+D^2(1-k\delta)^2}
\right]^2},
\nonumber\\
{\sf{b}}_2=
\frac{1}{2}
\sqrt{\left(\Omega_1-\Omega_2\right)^2
+4
\left[
\sqrt{I^2(1+\delta)^2+D^2(1+k\delta)^2}
-\sqrt{I^2(1-\delta)^2+D^2(1-k\delta)^2}
\right]^2},
\nonumber
\end{eqnarray}
and
${\mbox{E}}(\psi,a^2)\equiv
\int_0^{\psi}d\phi\sqrt{1-a^2\sin^2\phi}$
is the elliptic integral of the second kind.
We also seek for a nonzero solution 
$\delta^{\star}\ne 0$
of the equation
$\frac{\partial{\cal{E}}(\delta)}{\partial\delta}=0$.
Using (\ref{004})
one gets
\begin{eqnarray}
\label{005}
-\frac{1}{\pi}
{\mbox{E}}
\left(
\psi,
\frac{{\sf{b}}_1^2-{\sf{b}}_2^2}{{\sf{b}}_1^2}
\right)
\frac{\partial{\sf{b}}_1}{\partial\delta}
\nonumber\\
-\frac{1}{\pi}
\frac{{\sf{b}}_2^2
\frac{\partial{\sf{b}}_1}{\partial\delta}
-{\sf{b}}_1{\sf{b}}_2
\frac{\partial{\sf{b}}_2}{\partial\delta} 
}
{{\sf{b}}_1^2-{\sf{b}}_2^2}
\left[
{\mbox{E}}
\left(
\psi,
\frac{{\sf{b}}_1^2-{\sf{b}}_2^2}{{\sf{b}}_1^2}
\right)
-{\mbox{F}}
\left(
\psi,
\frac{{\sf{b}}_1^2-{\sf{b}}_2^2}{{\sf{b}}_1^2}
\right)
\right]
\nonumber\\
+2\alpha\delta=0,
\\
\frac{\partial{\sf{b}}_{1,2}}{\partial\delta}
=\frac{1}{{\sf{b}}_{1,2}}
\left[
\sqrt{I^2(1+\delta)^2+D^2(1+k\delta)^2}
\pm
\sqrt{I^2(1-\delta)^2+D^2(1-k\delta)^2}
\right]
\nonumber\\
\times
\left[
\frac{I^2(1+\delta)+kD^2(1+k\delta)}
{\sqrt{I^2(1+\delta)^2+D^2(1+k\delta)^2}}
\mp
\frac{I^2(1-\delta)+kD^2(1-k\delta)}
{\sqrt{I^2(1-\delta)^2+D^2(1-k\delta)^2}}
\right],
\nonumber
\end{eqnarray}
and ${\mbox{F}}(\psi,a^2)\equiv\int_0^{\psi}d\phi/
\sqrt{1-a^2\sin^2\phi}$
is the elliptic integral of the first kind.

Until the end of the paper we shall consider a case of the uniform 
transverse field
$\Omega_1=\Omega_2=\Omega_0$.
In the interesting for application limit
$\delta\ll 1$
valid for hard lattices having large values of $\alpha$
one finds
${\sf{b}}_{1}=2\vert{\cal{I}}\vert$,
${\sf{b}}_{2}=2\vert{\cal{I}}\vert\aleph\delta$
with
$\vert{\cal{I}}\vert=\sqrt{I^2+D^2}$
and
$\aleph=\frac{I^2+kD^2}{I^2+D^2}$
and instead of Eqs. (\ref{004}), (\ref{005})
one has
\begin{eqnarray}
\label{006}
e_0(\delta)=-\frac{2\vert{\cal{I}}\vert}{\pi}
{\mbox{E}}(\psi, 1-\aleph^2\delta^2)
-\vert\Omega_0\vert\left(\frac{1}{2}-\frac{\psi}{\pi}\right),
\\
\psi=
\left\{
\begin{array}{ll}
0 & 
{\mbox{if}}\;\;\;2\vert{\cal{I}}\vert<\vert\Omega_0\vert,\\
{\mbox{arcsin}}\sqrt{\frac{4{\cal{I}}^2-\Omega_0^2}
{4{\cal{I}}^2(1-\aleph^2\delta^2)}} &
{\mbox{if}}\;\;\;2\vert{\cal{I}}\vert\aleph\delta
\le\vert\Omega_0\vert<2\vert{\cal{I}}\vert,\\
\frac{\pi}{2} &
{\mbox{if}}\;\;\;
\vert\Omega_0\vert
<2\vert{\cal{I}}\vert\aleph\delta;
\end{array}
\right.
\nonumber
\end{eqnarray}
\begin{eqnarray}
\label{007}
\frac{\pi\alpha}{\vert{\cal{I}}\vert}
=\frac{\aleph^2}{1-\aleph^2\delta^2}
(F(\psi,1-\aleph^2\delta^2)
-E(\psi,1-\aleph^2\delta^2))
\end{eqnarray}

Consider the case $\Omega_0=0$.
After rescaling
${\cal{I}}\rightarrow I$,
$\frac{\alpha}{\aleph^2}\rightarrow\alpha$,
$\aleph\delta\rightarrow\delta$
one finds that 
Eq. (\ref{007})
is exactly as that considered in Ref. \cite{003} and thus
$\delta^{\star}\sim \frac{1}{\aleph}
\exp
\left(
-\frac{1}{\aleph^2}
\frac{\pi\alpha}{\vert{\cal{I}}\vert}
\right)$.
Thus 
for $k=1$ when $\aleph=1$
Dzyaloshinskii-Moriya interaction leads to increasing of
the dimerization parameter
$\delta^{\star}$, 
whereas for $k=0$ when $\aleph\le 1$
Dzyaloshinskii-Moriya interaction leads to decreasing of
the dimerization parameter
$\delta^{\star}$.

Consider further the case
$0<\vert\Omega_0\vert<2\vert{\cal{I}}\vert$.
Varying $\delta$ in the r.h.s. of Eq. (\ref{007}) from 0 to 1
one calculates a lattice parameter $\alpha$
for which the taken value of $\delta$ realizes an extremum of
${\cal{E}}(\delta)$.
One immediately observes that for
$\frac{\vert\Omega_0\vert}{2\vert{\cal{I}}\vert}\le
\aleph\delta$
the dependence
$\alpha$
versus
$\delta$
remains as that in the absence of the field,
whereas for
$0\le\aleph\delta<\frac{\vert\Omega_0\vert}{2\vert{\cal{I}}\vert}$
the calculated quantity
$\alpha$
starts to decrease.
From this one concludes that 
the field
$\frac{\vert\Omega_0\vert}{2\vert{\cal{I}}\vert}
=\exp\left(-\frac{1}{\aleph^2}
\frac{\pi\alpha}{\vert{\cal{I}}\vert}\right)$
makes the dimerization
unstable against the uniform phase.
The latter relation tells us that the Dzyaloshinskii-Moriya interaction
increases the value of that field for $k=1$ and decreases it for $k=0$.

In Figs. 1, 2 we presented the changes of the total energy 
${\cal{E}}(\delta)-{\cal{E}}(0)$
(\ref{004}) vs the dimerization parameter $\delta$ 
with switching on Dzyaloshinskii-Moriya interaction and the 
nonzero solution $\delta^{\star}$ 
of Eq. (\ref{005}) vs $\alpha$ with switching on Dzyaloshinskii-Moriya 
interaction, respectively. These results agree with the 
above ones valid in the 
limit $\delta\ll 1$.

Let us emphasize that one cannot treat 
rigorously
within the Jordan-Wigner picture 
complete Dzya\-lo\-shin\-skii-Moriya 
interaction that for neighbouring 
sites $n$ and $n+1$ reads
${\cal{D}}_n^x(s_n^ys_{n+1}^z-s_n^zs_{n+1}^y)
+{\cal{D}}_n^y(s_n^zs_{n+1}^x-s_n^xs_{n+1}^z)
+{\cal{D}}_n^z(s_n^xs_{n+1}^y-s_n^ys_{n+1}^x)$
except the latter term being included in
(\ref{001}).
The effects of ${\cal{D}}^x$, 
${\cal{D}}^y$ may be examined numerically performing 
finite-chain calculations.

To conclude, we have analysed a stability of
the spin-$\frac{1}{2}$
transverse $XX$ chain with respect to dimerization in the presence of
the Dzyaloshinskii-Moriya interaction
calculating for this purpose
with the help of continued fractions
the ground state energy for an arbitrary
value of the dimerization parameter.
Depending on the dependence of Dzyalosahinskii-Moriya interaction on the 
amplitude of lattice distortion it acts either in favour of dimerization or 
against it. 

It is generally known
\cite{002}
that the increasing of
the external field leads to a transition from the dimerized phase
to the incommensurate phase rather than to the uniform phase.
Evidently, the incommensurate phase cannot appear in the presented
treatment within the frames of the adopted ansatz for the
lattice distortions
$\delta_1\delta_2\delta_1\delta_2\ldots\;$,
$\delta_1+\delta_2=0$.
To clarify a possibility of more complicated distortions
the chains with longer periods should be examined.

\vspace{5mm}

The present study was partly supported by the DFG
(projects 436 UKR 17/20/98 and Ri 615/6-1).
O. D. acknowledges the kind hospitality of the Magdeburg University
in the spring
of 1999 when a part of the paper was done.
He is also indebted to Mrs. Olga Syska for continuous financial support.

\vspace{5mm}

FIGURE 1.
Dependence 
${\cal{E}}(\delta)-{\cal{E}}(0)$ vs $\delta$
for the spin-$\frac{1}{2}$ $XX$ chain with Dzyaloshinskii-Moriya 
interaction. 
$\vert I\vert=1$, 
$\vert \Omega_0\vert=0$,
$\alpha=0.8$,
a: $k=1$ 
($\vert D\vert=0,\;0.2,\;0.4,\;0.6,\;0.8,\;1$
from top to bottom), 
b: $k=0$ 
($\vert D\vert=0,\;0.2,\;0.4,\;0.6,\;0.8,\;1$
from bottom to top), 
c: $\vert D\vert=0.5$
($k=1,\;0.8,\;0.6,\;0.4,\;0.2,\;0$
from bottom to top). 

\vspace{5mm}
 
FIGURE 2.
Dependence $\delta^{\star}$ vs $\alpha$ 
for the spin-$\frac{1}{2}$ transverse 
$XX$ chain with Dzyaloshinskii-Moriya interaction. 
$\vert I\vert=1$,
$\alpha=0.8$,
$\Omega_0=0$ (a, d, g),
$\Omega_0=0.1$ (b, e, h),
$\Omega_0=0.3$ (c, f, i),
$k=1$ (a, b, c)
($\vert D\vert=0,\;0.2,\;0.4,\;0.6,\;0.8,\;1$
from left to right), 
$k=0$ (d, e, f) 
($\vert D\vert=0,\;0.2,\;0.4,\;0.6,\;0.8,\;1$
from right to left), 
$\vert D\vert=0.5$ (g, h, i)
($k=1,\;0.8,\;0.6,\;0.4,\;0.2,\;0$
from right to left). 

\clearpage

\begin{figure}[t]
\epsfysize=160mm
\epsfclipon
\centerline{\epsffile{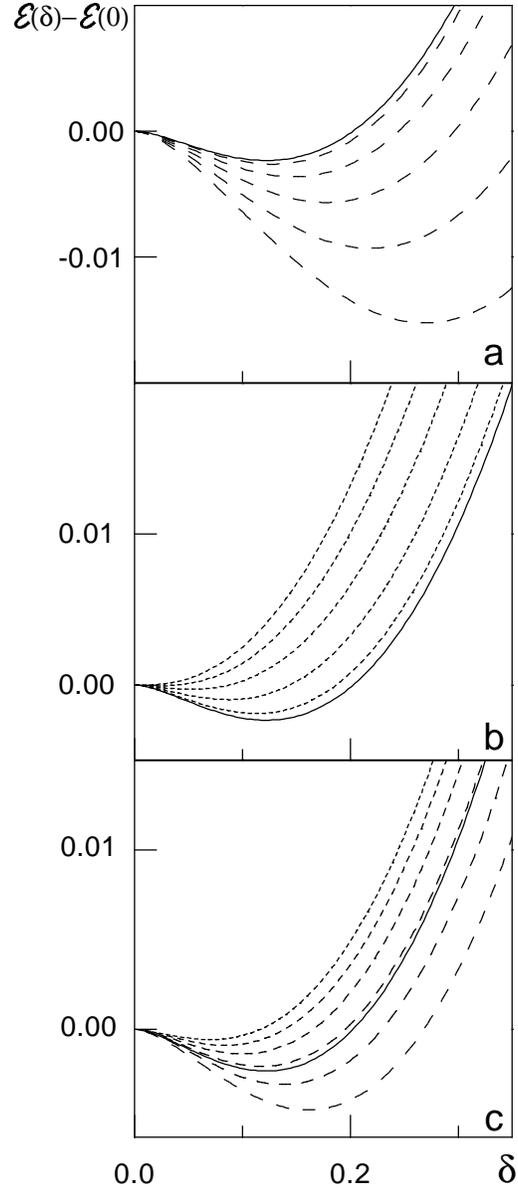}}
\caption[ ]
{Dependence 
${\cal{E}}(\delta)-{\cal{E}}(0)$ vs $\delta$
for the spin-$\frac{1}{2}$ $XX$ chain with Dzyaloshinskii-Moriya 
interaction. 
$\vert I\vert=1$, 
$\vert \Omega_0\vert=0$,
$\alpha=0.8$,
a: $k=1$ 
($\vert D\vert=0,\;0.2,\;0.4,\;0.6,\;0.8,\;1$
from top to bottom), 
b: $k=0$ 
($\vert D\vert=0,\;0.2,\;0.4,\;0.6,\;0.8,\;1$
from bottom to top), 
c: $\vert D\vert=0.5$
($k=1,\;0.8,\;0.6,\;0.4,\;0.2,\;0$
from bottom to top).}
\label{fig1}
\end{figure}

\clearpage

\begin{figure}[t]
\epsfysize=160mm
\epsfclipon
\centerline{\epsffile{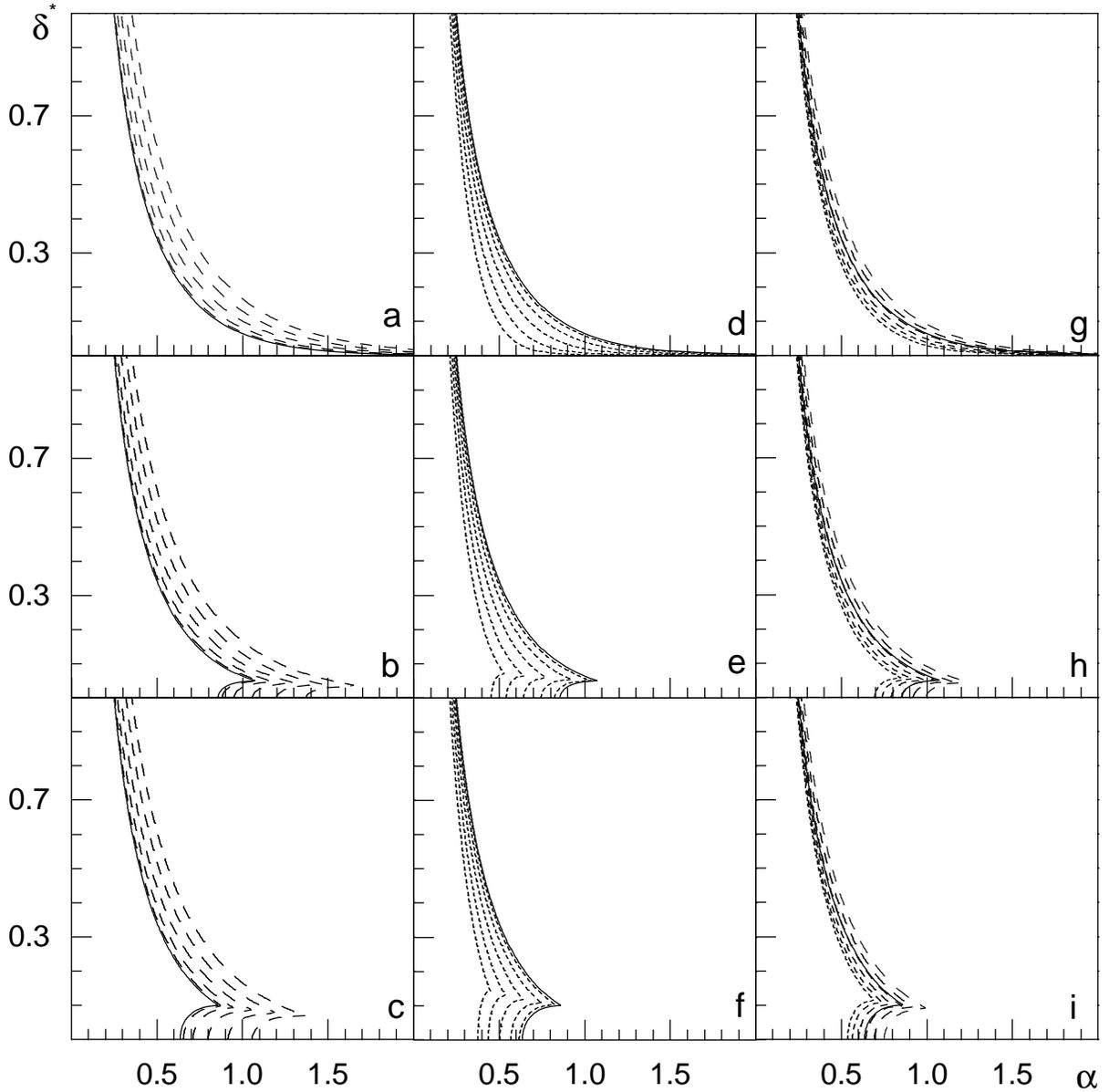}}
\caption[ ]
{Dependence $\delta^{\star}$ vs $\alpha$ 
for the spin-$\frac{1}{2}$ transverse 
$XX$ chain with Dzyaloshinskii-Moriya interaction. 
$\vert I\vert=1$,
$\alpha=0.8$,
$\Omega_0=0$ (a, d, g),
$\Omega_0=0.1$ (b, e, h),
$\Omega_0=0.3$ (c, f, i),
$k=1$ (a, b, c)
($\vert D\vert=0,\;0.2,\;0.4,\;0.6,\;0.8,\;1$
from left to right), 
$k=0$ (d, e, f) 
($\vert D\vert=0,\;0.2,\;0.4,\;0.6,\;0.8,\;1$
from right to left), 
$\vert D\vert=0.5$ (g, h, i)
($k=1,\;0.8,\;0.6,\;0.4,\;0.2,\;0$
from right to left).}
 \label{fig1}
\end{figure}

\end{document}